\documentstyle[12pt]{article}

\newcommand{\ba}{\begin{eqnarray}}
\newcommand{\ea}{\end{eqnarray}}
\begin{document}
\pagestyle{plain}
\title{Spectrum generating algebra for
X$_{3}$ molecules}
\author{R.~Bijker$^{a,b}$, A.E.L.~Dieperink$^{c}$ and A.~Leviatan$^{d}$\\
\and
\begin{tabular}{rl}
$^{a}$&R.J.~Van de Graaff Laboratory, University of Utrecht,\\
      &P.O. Box 80000, 3508 TA Utrecht, The Netherlands\\
$^{b}$&Instituto de Ciencias Nucleares, U.N.A.M.,\\
      &A.P. 70-543, 04510 M\'exico D.F., M\'exico \thanks{Present address}\\
$^{c}$&K.V.I., Zernikelaan 25, 9747 AA Groningen, The Netherlands\\
$^{d}$&Racah Institute of Physics, The Hebrew University,\\
      &Jerusalem 91904, Israel
\end{tabular}}
\date{}
\maketitle
\noindent
\vspace{6pt}
\begin{abstract}
A new spectrum generating algebra for a unified description of
rotations and vibrations in polyatomic molecules is introduced.
An application to nonlinear X$_3$ molecules shows that this model
(i) incorporates exactly the relevant point group, (ii) provides a complete
classification of oblate top states, and (iii) treats properly
both degenerate and nondegenerate vibrations.
\end{abstract}
%\noindent
\newpage
\section{Introduction}

The progress currently witnessed in experimental techniques for the
spectroscopy of large molecules motivates the development of new theoretical
tools to interpret and guide such measurements. In view of the
proliferation of parameters needed in Dunham expansions and the difficulties
encountered in solving a Schr\"odinger equation with interatomic potentials
for polyatomic molecules, it has been suggested to use algebraic methods
in which the hamiltonian is expressed in terms of elements of a Lie algebra.
The key ingredient is the choice of a suitable spectrum generating algebra.

At present there exist two versions of this method. In the first version
\cite{vibron1,vibron2}, rotations and vibrations are treated simultaneously,
and hence rotation-vibration couplings are built in from the outset
\cite{rotvib}. In this approach, a molecule with $n$ atoms is described
in terms of $n-1$ coupled $U(4)$ groups, one for each independent relative
coordinate. Each $U_{i}(4)$ algebra is realized in terms of a set of four
vibron operators ($\sigma_i$, $\pi_i$): the three components of
a dipole (or $\pi$-) boson with $L^P=1^-$ and a scalar (or $\sigma$-)
boson with $L^P=0^+$, {\it i.e.} one $\sigma$-boson for every $\pi$-boson.
The scalar boson does not represent an independent degree of freedom, but
is merely introduced to conveniently handle anharmonicities by
compactifying the model space, {\it i.e.}
the number of bosons $N_i= n_{\sigma_i}+n_{\pi_i}$, is conserved for
each $i=1,\ldots,n-1$ separately.
This version has been applied successfully to polyatomic molecules
with $n=2,3,4$ atoms, but it encounters difficulties in describing bent
molecules with degenerate vibrations, such as for example
in nonlinear X$_3$ molecules \cite{vibron2,BDL}.

In the second version \cite{vibronu2a,vibronu2b} rotations are ignored,
while vibrations are treated in
terms of coupled one-dimensional anharmonic oscillators. A separate set of
coupled $U(2)$ groups is introduced for each type of vibration (stretching
and bending). The use of symmetry adapted operators ensures the correct
transformation properties of the eigenstates under the relevant point
group \cite{vibronu2b}. Since there is no explicit coupling between
rotations and vibrations, this scheme becomes particularly useful for
molecules, in which the rotation-vibration coupling is negligible.
In this approach each type of vibration is treated separately with
different interactions, although for degenerate vibrations, such as
for example in symmetric top X$_3$ molecules, stretching and bending
vibrations can belong to the same irreducible representation (irrep) of
the relevant point group.

In this note, we propose an alternative scheme for the description
of $n$-atomic molecules, in which the relevant point group symmetry
is taken into account {\em exactly} and
which contains {\em all} vibrations (stretching and bending) and
rotations in a single algebraic framework.
We introduce a dipole boson for each independent relative coordinate,
and a single scalar boson.
This leads to a spectrum generating algebra of $U(k+1)$, where $k=3(n-1)$
is the total number of rotational and vibrational degrees of freedom.
By construction, only the total number of bosons,
$N= n_{\sigma}+n_{\pi}$ with $n_{\pi}=\sum_i n_{\pi_i}$ is conserved.
For diatomic molecules ($n=2$) we recover the $U(4)$ vibron model
\cite{vibron1}. For triatomic molecules ($n=3$)
we obtain a $U(7)$ model whose
building blocks are a scalar boson and two dipole bosons
($\sigma$, $\pi_{1}$, $\pi_{2}$).
We present this scheme by studying symmetric (oblate) top
X$_{3}$ molecules which form the simplest nontrivial example of a
bent polyatomic molecule with a degenerate vibration.
We compare our results with those obtained in a $U(4) \times U(4)$
model \cite{vibron2}.

\section{Point group symmetry}

The embedding of discrete point group symmetries in an algebraic model
relies on a geometric interpretation of the vibrons.
For nonlinear rigid molecules it is convenient
to make use of the established isomorphism between the
molecular symmetry group and the relevant point group \cite{bunker}. The
elements of the former consist of permutations of identical nuclei with
or without inversion (parity). The parity of the bosons is well defined
and the transformation properties under permutations can be realized
in terms of finite rotations among the bosons.
For bent X$_{3}$ molecules with point group symmetry $D_{3h}$, we use the
isomorphism between the $S_{3}$ permutation group and the point group
$D_{3}$ ($\subset D_{3h}$). The parity and $S_3$ labels are equivalent to
the classification under $D_{3h}$. We associate the two types of dipole
bosons, $\pi_{1}$ and $\pi_{2}$, with the relative Jacobi coordinates
\ba
\vec{\rho} \;=\; \frac{1}{\sqrt{2}}
(\vec{x}_1 - \vec{x}_2) ~, \hspace{1cm}
\vec{\lambda} \;=\; \frac{1}{\sqrt{6}}
(\vec{x}_1 + \vec{x}_2 - 2\vec{x}_3) ~, \label{jacobi}
\ea
and their conjugate momenta, which have well defined transformation
properties under permutations ($\vec{x}_i$ denotes the coordinate of the
$i$-th atom). The elements of the group $S_3$ can be expressed in terms
of the transposition $P(12)$ and the cyclic permutation $P(123)$ whose
matrix representation in the ($\sigma^{\dagger}$, $\pi^{\dagger}_{1,m}$,
$\pi^{\dagger}_{2,m}$) basis is \cite{MM}
\ba
P(12) \;=\; \left( \begin{array}{rrr} 1 & 0 & 0 \\ 0 & -1 & 0 \\
0 & 0 & 1 \end{array} \right) ~, \hspace{1cm}
P(123) \;=\; \left( \begin{array}{ccc} 1 & 0 & 0 \\
0 & -1/2 &  \sqrt{3}/2 \\ 0 & -\sqrt{3}/2 & -1/2 \end{array} \right) ~.
\ea
The $\sigma$-boson is a scalar under the permutation group, whereas
the two $\pi$-bosons transform (for each projection $m=0,\pm 1$) as the
two components ($M_{\rho}$, $M_{\lambda}$) of the two-dimensional irrep
($M$) of $S_3$. In general, the symmetric ($S$), antisymmetric ($A$),
and mixed symmetry ($M$) classes of $S_3$, can equivalently be labeled by
the irreps of the isomorphic point group $D_{3}$, as $A_{1}$, $A_2$ and $E$,
respectively.

The transformation properties under $S_3$ of all operators of interest
follow from those of the building blocks. In particular, the most general
one- and two-body $U(7)$ hamiltonian which is a scalar under $S_3$
as well as rotationally and parity invariant is found to be
\ba
H &=& \epsilon_{\sigma} \, \sigma^{\dagger} \sigma
- \epsilon_{\pi} \,
(\pi_{1}^{\dagger} \cdot \tilde{\pi}_{1}
+\pi_{2}^{\dagger} \cdot \tilde{\pi}_{2})
+ u_0 \, \sigma^{\dagger} \sigma^{\dagger} \sigma \sigma - u_1 \,
  \sigma^{\dagger} ( \pi^{\dagger}_{1} \cdot \tilde{\pi}_{1}
+ \pi^{\dagger}_{2} \cdot \tilde{\pi}_{2} ) \sigma
\nonumber\\
&& +\, v_0 \, \left[ ( \pi^{\dagger}_{1} \cdot \pi^{\dagger}_{1}
  + \pi^{\dagger}_{2} \cdot \pi^{\dagger}_{2} ) \sigma \sigma
  + \sigma^{\dagger} \sigma^{\dagger}
  ( \tilde{\pi}_{1} \cdot \tilde{\pi}_{1}
  + \tilde{\pi}_{2} \cdot \tilde{\pi}_{2} ) \right]
\nonumber\\
&& + \sum_{\lambda=0,2} c_{\lambda} \,
\left[ ( \pi^{\dagger}_{1} \pi^{\dagger}_{1}
  - \pi^{\dagger}_{2} \pi^{\dagger}_{2} )^{(\lambda)} \cdot
  ( \tilde{\pi}_{1} \tilde{\pi}_{1}
  - \tilde{\pi}_{2} \tilde{\pi}_{2} )^{(\lambda)}
+ 4 \, ( \pi^{\dagger}_{1} \pi^{\dagger}_{2})^{(\lambda)} \cdot
       ( \tilde \pi_{2} \tilde \pi_{1})^{(\lambda)} \right]
\nonumber\\
&& + c_1 \, ( \pi^{\dagger}_{1} \pi^{\dagger}_{2} )^{(1)} \cdot
( \tilde \pi_{2} \tilde \pi_{1} )^{(1)}
+ \sum_{\lambda=0,2} w_{\lambda} \,
  ( \pi^{\dagger}_{1} \pi^{\dagger}_{1}
  + \pi^{\dagger}_{2} \pi^{\dagger}_{2} )^{(\lambda)} \cdot
  ( \tilde{\pi}_{1} \tilde{\pi}_{1}
  + \tilde{\pi}_{2} \tilde{\pi}_{2} )^{(\lambda)} ~,\quad\;\;
\label{hs3}
\ea
with $\tilde\pi_{i,m}=(-)^{1-m}\pi_{i,-m}$ and $i=1,2$.
The corresponding eigenstates are labeled by the total number of
bosons $N$ and, by construction, have good angular momentum,
parity and permutation (point group) symmetry.

\section{Geometry}

The hamiltonian of Eq.~(\ref{hs3}) is expressed in terms of abstract
algebraic interactions. A more intuitive geometric visualization can be
obtained by using mean-field techniques to study the geometric properties
of the $U(7)$ model. For a system of bosons the variational wave function
takes the form of a coherent state \cite{ROOS} which is a condensate of
$N$ bosons,
\ba
\mid N;c \rangle &=&
\frac{1}{\sqrt{N!}} \Bigl( b_c^{\dagger} \Bigr)^N \mid 0 \rangle ~,
\label{cond}
\ea
with
\ba
b_c^{\dagger} &=& (1+R^2)^{-1/2}
\Bigl[ \sigma^{\dagger} + r_{1} \, \pi_{1, z}^{\dagger}
+ r_{2} \, ( \cos \theta \, \pi_{2, z}^{\dagger}
         + \sin \theta \, \pi_{2, x}^{\dagger} ) \Bigr] ~,
\label{bc}
\ea
where $R^{2} = r_{1}^{2} + r_{2}^{2}$. The condensate is parametrized
in terms of two (real) coordinates, $r_{1}$ and $r_{2}$, and an angle
$\theta$, ($r_{1},r_{2} \geq 0$ and $0 \leq \theta \leq \pi$).
The two vectors $\vec{r}_{1}$ and $\vec{r}_{2}$ span the $xz$-plane.
We have chosen the $z$-axis along the direction of $\vec{r}_{1}$,
and $\vec{r}_{2}$ is rotated by an angle $\theta$ about the out-of-plane
$y$-axis, $\vec{r}_{1} \cdot \vec{r}_{2} = r_{1} r_{2} \cos \theta$.
The expectation value of the hamiltonian of Eq.~(\ref{hs3})
in the condensate defines a classical energy surface
$E(r_{1}, r_{2}, \theta)$. The equilibrium shape is determined by minimizing
the energy surface with respect to $r_{1}$, $r_{2}$ and
$\theta$. The nonlinear rigid and stable equilibrium shape
is characterized by $r_{1}=r_{2}$ and $\theta = \pi/2$.
These two conditions are precisely
those satisfied by the Jacobi coordinates
of Eq.~(\ref{jacobi}) for an equilateral triangular shape,
and suggests to associate the algebraic shape parameters in
Eq.~(\ref{bc}) with these coordinates, {\it i.e.}
$r_{1}\leftrightarrow \rho$ and $r_{2}\leftrightarrow \lambda$~.

\section{Excitations}

We consider separately the vibrational and rotational
excitations described by the hamiltonian in Eq.~(\ref{hs3}). With the
techniques introduced in \cite{KL} for the nuclear interacting boson
model, an arbitrary $U(7)$ hamiltonian can be decomposed uniquely into
intrinsic (vibrational) and collective (rotational and rotation-vibration
coupling) parts,
\ba
H &=& H_{\mbox{int}} + H_{\mbox{coll}} ~.
\ea

\subsection{Vibrations}

The intrinsic part by definition annihilates the equilibrium condensate
and has the same shape for the energy surface as the original hamiltonian.
For the rigid triangular equilibrium shape, characterized by $r_{1}=r_{2}$
and $\theta = \pi/2$, we find
\ba
H_{\mbox{int}} &=& \xi_1 \,
\Bigl ( R^2 \, \sigma^{\dagger} \sigma^{\dagger}
- \pi^{\dagger}_{1} \cdot \pi^{\dagger}_{1}
- \pi^{\dagger}_{2} \cdot \pi^{\dagger}_{2} \Bigr ) \,
\Bigl ( R^2 \, \sigma \sigma - \tilde{\pi}_{1} \cdot \tilde{\pi}_{1}
- \tilde{\pi}_{2} \cdot \tilde{\pi}_{2} \Bigr )
\nonumber\\
&& + \xi_2 \, \Bigl [
\Bigl ( \pi^{\dagger}_{1} \cdot \pi^{\dagger}_{1}
- \pi^{\dagger}_{2} \cdot \pi^{\dagger}_{2} \Bigr ) \,
\Bigl ( \tilde{\pi}_{1} \cdot \tilde{\pi}_{1}
- \tilde{\pi}_{2} \cdot \tilde{\pi}_{2} \Bigr )
+ 4 \, \Bigl ( \pi^{\dagger}_{1} \cdot \pi^{\dagger}_{2} \Bigr ) \,
\Bigl ( \tilde \pi_{2} \cdot \tilde \pi_{1} \Bigr ) \Bigr ] ~.\qquad
\label{hint}
\ea
%Characteristics of the vibrational spectrum can be studied by introducing
To get more insight into the vibrational structure, we perform a
normal-mode analysis to the above hamiltonian. This is done by introducing
a set of orthonormal deformed vibrons, consisting of the condensate boson
of Eq.~(\ref{bc}) with $r_{1}=r_{2}$ and $\theta=\pi/2$, and six additional
bosons which represent excitations of the condensate.
The normal modes can be found by rewriting the intrinsic hamiltonian
in terms of the deformed vibrons and replacing the condensate bosons by
their classical mean-field value $\sqrt{N}$ \cite{LK,point}. As a result
we find to leading order in $N$
\ba
\frac{1}{N} H_{\mbox{int}} &=& \epsilon_1 \, b_u^{\dagger} b_u
+ \epsilon_{2} \, (b_v^{\dagger} b_v + b_w^{\dagger} b_w)
+ {\cal O}(1/\sqrt{N}) ~, \label{vint}
\ea
with eigenfrequencies $\epsilon_1 = 4 \xi_1 R^2$ and
$\epsilon_2 = 4 \xi_2 R^2 (1+R^2)^{-1}$.
This identifies the deformed bosons that correspond to the
three fundamental vibrations: a symmetric stretching ($u$),
an antisymmetric stretching ($v$) and a bending vibration ($w$),
%\vfil\break
\ba
b^{\dagger}_{u} &=& (1+R^2)^{-1/2} \Bigl[ -R \, \sigma^{\dagger} +
\Bigl( \pi_{1, z}^{\dagger}+\pi_{2, x}^{\dagger} \Bigr)/\sqrt{2}\, \Bigr] ~,
\nonumber\\
b^{\dagger}_{v} &=&
\Bigl( -\pi_{1, z}^{\dagger}+\pi_{2, x}^{\dagger} \Bigr)/\sqrt{2} ~,
\nonumber\\
b^{\dagger}_{w} &=&
\Bigl(\pi_{1, x}^{\dagger}+\pi_{2, z}^{\dagger} \Bigr)/\sqrt{2} ~.
\label{bi}
\ea
The first two are radial excitations, whereas the third is an angular
mode which corresponds to oscillations in the angle $\theta$ between the
two Jacobi coordinates \cite{point} ($\nu_1$, $\nu_{2a}$ and $\nu_{2b}$,
respectively in the usual spectroscopic notation).
The angular mode is degenerate with the antisymmetric radial mode.
This is in agreement with the point group classification of the
fundamental vibrations for a symmetric X$_3$ configuration
\cite{Herzberg} and shows that $H_{\mbox{int}}$ describes the vibrational
excitations of an oblate symmetric top. Intrinsic states representing
excited vibrations $(\nu_1,\nu_2^{l})$, are obtained, for large $N$, by
replacing the condensate bosons in Eq.~(\ref{cond})
by the appropriate number of deformed bosons of Eq.~(\ref{bi}).
Accordingly, the intrinsic state for such vibration (having $+l$
projection of the vibrational angular momentum on the symmetry $y$-axis)
takes the form $(b^{\dagger}_{\theta,1})^{(\nu_2+l)/2}
(b^{\dagger}_{\theta,-1})^{(\nu_2-l)/2}
(b^{\dagger}_{u})^{\nu_1}
(b^{\dagger}_{c})^{N-\nu_1-\nu_2}\mid 0 \rangle$, where
$b^{\dagger}_{\theta,\pm 1} = (b^{\dagger}_{v}\pm i
b^{\dagger}_{w})/\sqrt{2}$.

By construction, $H_{\mbox{int}}$ has an exactly degenerate ground band
whose rotational members are obtained by projection from the
equilibrium condensate. Its excited states also tend to cluster into bands.
Eq.~(\ref{vint}) shows that in the large $N$ limit the vibrational
spectrum is harmonic. For finite values of $N$ the correction terms of order
${\cal O}(1/\sqrt{N})$ give rise to anharmonicities. These can be
studied numerically by diagonalizing $H_{\mbox{int}}$ in a convenient
basis.

A similar analysis \cite{point} of the $S_3$-invariant intrinsic (one- and
two-body) hamiltonian of the $U(4) \otimes U(4)$ model yields that,
in this case, the two radial modes associated with the Jacobi
coordinates are uncoupled and have the same frequency,
in disagreement with the point group classification of the normal
vibrations of a symmetric X$_3$ shape.
The difference with the $U(7)$ model can be traced back to the occurrence
of terms in the hamiltonian of Eq.~(\ref{hint}) in which only the
total number of dipole bosons is conserved. These terms cannot appear
in a $U(4) \otimes U(4)$ description which requires that each type of bosons
is conserved separately.
A $U(2) \otimes U(2)$ description will yield only the symmetric
$\nu_{1}(A_{1})$ and the antisymmetric $\nu_{2a}(E)$ stretching vibrations.
The second member of the degenerate vibration (the $\nu_{2b}(E)$ bending)
requires a separate treatment.

\subsection{Rotations}

On top of each vibrational excitation there is a whole series of
rotational states. In a geometric description the rotational
excitations are labeled by the angular momentum $L$ and its projection
$K$ ($=0,1,\ldots$) on the threefold symmetry axis, parity $P=(-)^{K}$
and the transformation property $t$ under the point group.
In the present algebraic model the rotations (and rotation-vibration
couplings) are described by the collective part of the hamiltonian.
By construction, $H_{\mbox{coll}}=H-H_{\mbox{int}}$ consists
of interaction terms which do not affect the shape of the energy surface
\cite{KL,LK}. Discarding $\hat{N}$-dependent terms that do not
contribute to the excitation spectrum, we find
\ba
H_{\mbox{coll}} &=&
 \kappa_1 \, ( \hat A_1 \cdot \hat A_1 + \hat A_2 \cdot \hat A_2 )
+ \kappa_2 \, ( \hat B_1 \cdot \hat B_1 + \hat B_2 \cdot \hat B_2 )
\nonumber\\
&&
+\, \kappa_3 \, \hat L \cdot \hat L
+ \kappa_4 \, \hat K_y \cdot \hat K_y ~,
\label{hcoll}
\ea
with
\ba
\begin{array}{ll}
\hat A_1 = i \, (\pi^{\dagger}_1 \sigma +
\sigma^{\dagger} \tilde{\pi}_1)^{(1)} ~, &
\hat A_2 = i \, (\pi^{\dagger}_2 \sigma +
\sigma^{\dagger} \tilde{\pi}_2)^{(1)} ~, \\
\hat B_1 = ( \pi^{\dagger}_1 \tilde{\pi}_2
+ \pi^{\dagger}_2 \tilde{\pi}_1 )^{(1)} ~, &
\hat B_2 = ( \pi^{\dagger}_1 \tilde{\pi}_1
- \pi^{\dagger}_2 \tilde{\pi}_2 )^{(1)} ~, \\
\hat L = \sqrt{2} \, ( \pi^{\dagger}_1 \tilde{\pi}_1
+ \pi^{\dagger}_2 \tilde{\pi}_2 )^{(1)} ~, &
\hat K_y = -i \sqrt{3} \, ( \pi^{\dagger}_1 \tilde{\pi}_2 -
\pi^{\dagger}_2 \tilde{\pi}_1 )^{(0)} ~.
\end{array}
\ea
The angular momentum $\hat L$ commutes with any rotational-invariant
hamiltonian. The hamiltonian of Eq.~(\ref{hs3}) commutes also with
the operator $\hat K_y$. Consequently the resulting eigenstates
$\vert L^{P},K_y\,\rangle$ have good angular momentum $L$,
parity $P$ and $K_y$.
The states $\vert L^{P},\pm K_y\,\rangle$ are degenerate since
$P(12)\, \vert L^{P},K_y\,\rangle \,=\,(-)^{K_y}\,\vert L^{P},-K_y\,\rangle $
and $P(12)$ commutes with any $S_3$ invariant hamiltonian.
The operator $\hat K_y$ is related to the cyclic permutation
$P(123)=\exp(-i2\pi\hat K_y/3)$ and the absolute value $|K_y|$ determines
the $S_3\approx D_3$ symmetry \cite{BOW},
\ba
|K_y| &=& \left\{ \begin{array}{rl}
0   \mbox{ (mod 3)} & \hspace{1cm} \mbox{for } A_1, A_2 \\
1,2 \mbox{ (mod 3)} & \hspace{1cm} \mbox{for } E
\end{array} \right.
\ea
States with well defined permutation symmetry ($t$) can be formed by
taking the two linear combinations
$\vert L^{P}_t, |K_y|\,\rangle \propto\Bigl [ 1 \pm P(12)\Bigr ]
\vert L^{P},K_y\,\rangle $, which can be used to distinguish between
$t=A_1$ and $A_2$ for $|K_y|=0$ (mod 3), and between the components
$E_{\rho}$ and $E_{\lambda}$ for $|K_y|=1,2$ (mod 3).

By examining the projection of intrinsic states on the symmetry axis,
it is possible to show that for a $(\nu_1,\nu_2^{l})$ vibration,
the algebraic quantum number $|K_y|$ and the geometric $K$ are related by
\ba
|K_y|\,=\, |K \pm 2l| ~. \label{ky}
\ea
The $+(-)$ sign corresponds to the $+l(-l)$ projections of the
vibrational angular momentum along the symmetry axis.
The $|K_y|$ quantum number is analogous to the $G$ quantum number
\ba
G\,=\, |K \pm l| ~, \label{gwatson}
\ea
considered by Watson \cite{wat}.
Since both the $G$ and $|K_y|$ labels are defined (mod 3), they
provide an equivalent classification scheme.
It is important to note that, unlike for $l=0$
where $|K_y|=G=|K|$, for $l>0$ there are two possible $|K_y|$ or $G$
values for each $K$. Thus $|K_y|$ (or $G$) is an additional quantum number
needed to supplement the $D_3$ and parity labels for a complete
classification of the oblate top states. As an example,
in $E$-vibrations of the type $(\nu_1,\nu_2^{l=1})$
the two $L^{-}_{E}$ levels with $K=3$ are
distinguished by $|K_y|=5$ $(G=2)$ and $|K_y|=1$ $(G=4)$.

{}From the above discussion it is evident that the last two terms in
Eq.~(\ref{hcoll}) commute with the hamiltonian of Eq.~(\ref{hs3})
and thus correspond to {\em exact} symmetries. Their
eigenvalues $\kappa_{3} \, L(L+1) + \kappa_{4} \, K^2_y$,
are similar in form to those of a symmetric top.
All rotational states with $K\neq 0$ are split in two, with
different $|K_y|$ assignments in accord with Eqs. (12) and (13).
%~(\ref{ky}) and~(\ref{gwatson}).
The splitting between $+l$ and $-l$ levels is $8\kappa_4lK$, {\it i.e.}
increases linearly with $K$. The $\kappa_1$ and $\kappa_2$ terms in
Eq.~(\ref{hcoll}) do not commute with the intrinsic hamiltonian of
Eq.~(\ref{hint}).
Therefore, in addition to shifting and splitting the bands generated
by $H_{\mbox{int}}$, they can also mix them and hence contain the
rotation-vibration couplings. Their effect on
the spectrum can be studied numerically.

The one- and two-body $U(7)$ hamiltonian presented so far
does not exhibit $l$-type doubling
and consequently its calculated $A_1$ and $A_2$ levels occur
in degenerate doublets.
This degeneracy can be lifted with the inclusion
of higher-order terms which break the $|K_y|$ symmetry.
An example of such a three-body term is considered below.
A similar situation is encountered in
Watson's effective hamiltonian \cite{wat}, whose major terms are
diagonal in the quantum number $G$, and small higher-order correction
terms  (with $\Delta G=6$) are important for the splitting of the
$(A_1,A_2)$ doublets.

\subsection{Applications}

Among the new features in the proposed $U(7)$ spectrum generating
algebra, as compared to the $U_{1}(4)\otimes U_{2}(4)$ scheme,
is the exact account of the point symmetry. This is particularly
important for degenerate vibrations present in symmetric top X$_3$
molecules. With that in mind,
we apply the present formalism to the vibrational spectrum of
H$_{3}^{+}$. This molecule, with a triangular equilibrium shape,
has been studied extensively \cite{oka} and can serve as a test ground
for different spectrum generating algebras. It was shown \cite{BDL,point}
that the $U(4) \otimes U(4)$ model encounters some difficulties
in the description of the vibrational spectrum of
H$_{3}^{+}$ \cite{vibron2}. Our goal here is to examine to what
extent these difficulties arise from the lack of an exact treatment
of the molecular point group symmetry.
For that purpose we use a simple $U(7)$ hamiltonian
\ba
H &=& H_{\mbox{int}}(\xi_1,\xi_2,R^2)
+ \kappa_4\hat K_y\cdot\hat K_y
+\zeta \,T^{\dagger} \cdot \tilde T  ~,
\ea
where  $H_{\mbox{int}}$ is given in Eq. (7) and
$T^{\dagger} = (\pi^{\dagger}_1\cdot\pi^{\dagger}_1 -
\pi^{\dagger}_2\cdot\pi^{\dagger}_2)\pi^{\dagger}_2 +
2(\pi^{\dagger}_1\cdot\pi^{\dagger}_2)\pi^{\dagger}_1$.
The parameters are determined from a fit
to the estimated band origins in H$_{3}^{+}$ \cite{dmt}.
In the fit we minimize $\sum_i w_i \, [ E_i(\mbox{calc})-E_i ]^2$
with weights $w_i=1.0$ for the two fundamentals (the only band origins
known from experiment) and $w_i=0.1$ for the other levels.
As shown in Table 1, the $U(7)$ calculation shows a significant
improvement over the $U_{1}(4)\otimes U_{2}(4)$ fit, thus
highlighting the importance of incorporating the discrete point group
symmetry in the algebraic description.
It is clear, however, that the simple $U(7)$ hamiltonian (15) used here
is not sufficient to reach the accuracy of ab initio
calculations. Adding more interactions
({\it e.g.} rotation-vibration terms
and/or additional higher-order terms in the algebraic hamiltonian)
can improve the spectroscopic accuracy of the fit at the expense of more
parameters. The need for such terms is expected in view of
the known large rotation-vibration couplings in H$_{3}^{+}$,
the non-Born Oppenheimer corrections in the H$_{3}^{+}$ potential
\cite{tenn} and the presence of higher-order terms in ab initio
potential surfaces (31 terms \cite{dmt}).

\section{Summary and conclusions}

In this work we have introduced a $U(7)$ spectrum generating algebra
for the description of triatomic molecules.
Particular emphasis was put on the ability of the model to adequately
treat degenerate vibrations as they occur in oblate top X$_{3}$ molecules,
by imposing the discrete point group symmetry on the hamiltonian.
A normal mode analysis revealed that the characteristic pattern of
fundamental vibrations of a symmetric bent X$_3$ molecule is recovered
in this model. Algebraic terms affecting the vibrations, rotations
and their coupling have been identified.
The $U(7)$ algebra provides a quantum number $|K_y|$ (the analog
of the quantum number $G$), needed
for a complete classification of oblate top states.
Although we have focussed the discussion to oblate top X$_{3}$ molecules,
the $U(7)$ model can be applied to any triatomic molecule, both linear and
bent. In each case the results of a normal mode analysis are in agreement
with those dictated by the relevant point group \cite{BDL}.
All these features of the $U(7)$ model are necessary ingredients for a
proper description of rotations and vibrations in triatomic molecules.

We have compared the present model with two other algebraic approaches.
The main difference between the models
lies in the choice of the model space. In the $U(7)$ model the
distribution of quanta among the two dipole degrees of freedom is
determined dynamically by
the hamiltonian, whereas in the $U(4) \otimes U(4)$
model (or its simplified $U(2) \otimes U(2)$ version)
the number of bosons in each mode is restricted separately.
The existence of generators in the $U(7)$ algebra that mix the two
types of dipole bosons, is the new ingredient that enables an exact
treatment of the point group symmetry and a proper description
of degenerate vibrations. Unlike in the $U_{1}(4)\otimes U_{2}(4)$ case,
in the $U(7)$ scheme there is no need to use non-analytic (absolute value)
interaction terms to describe bent molecules.
On the other hand the $U(7)$ algebra does not have a
$SO(4)\otimes SO(4)$ dynamical symmetry, which led to considerable
simplification in the $U(4)\otimes U(4)$ description
of linear molecules. In a fit to the band origins of H$_{3}^{+}$ we showed
that a simple $U(7)$ hamiltonian with the proper point group symmetry
improves considerably the previous $U_{1}(4)\otimes U_{2}(4)$ description
of triangular $X_3$ molecules.

Refinements of the simple $U(7)$ hamiltonian are needed to achieve
spectroscopic accuracy. The choice of rotation-vibration terms and of
additional higher-order algebraic terms may be guided by examining
corresponding terms in Watson's effective hamiltonian. The inherent
simplicity in algebraic methods is potentially important for polyatomic
molecules, where are ab initio calculations are difficult to
perform. A first step in this direction would be to extend
the current procedure to larger molecules, resulting in a spectrum
generating algebra of $U(3n-2)$ as a candidate
for a spectrum generating algebra for
vibrations and rotations in $n$-atomic molecules
with prescribed point group symmetries.
Work along these lines, as well as on the rotational aspects of these
models, is in progress and will be reported separately.

\section*{Acknowledgements}

This work is supported in part by the Stichting voor Fundamenteel
Onderzoek der Materie (FOM) with financial support from the
Nederlandse Organisatie voor Wetenschappelijk Onderzoek (NWO),
by CONACyT, M\'exico under project
400340-5-3401E, DGAPA-UNAM under project IN105194,
and the Israeli Science Ministry.

\clearpage
\begin{table}
\centering
\caption{H$_{3}^{+}$ band origins in cm$^{-1}$.}
\vspace{10pt}
\begin{tabular}{cc|c|c|cc}
\hline
& & & & & \\
($\nu_1,\nu_{2}^{l}$) & $t$ &
DMT$^{a)}$ & $U(7)^{b)}$ & \multicolumn{2}{c}{$U(4) \otimes U(4)^{c)}$}\\
& & & & I & II \\
& & & & & \\
\hline
& & & & & \\
($0,0^{0}$) & $A_1$ & 0    & 0    & 0    & 0    \\
($0,1^{1}$) & $E  $ & 2521 & 2522 & 2506 & 2506 \\
($1,0^{0}$) & $A_1$ & 3178 & 3184 & 3181 & 3174 \\
($0,2^{0}$) & $A_1$ & 4778 & 4770 & 4928 & 4929 \\
($0,2^{2}$) & $E$   & 4998 & 5007 & 4928 & 4929 \\
($1,1^{1}$) & $E$   & 5554 & 5543 & 5603 & 5597 \\
($2,0^{0}$) & $A_1$ & 6262 & 6233 & 6233 & 6243 \\
($0,3^{1}$) & $E$   & 7006 & 7019 & 7266 & 7269 \\
($0,3^{3}$) & $A_1$ & 7283 & 7276 & 7266 & 7269 \\
($0,3^{3}$) & $A_2$ & 7493 & 7492 & 7266 & 7269 \\
& & & & & \\
\hline
\multicolumn{6}{l}{$^{a)}$ From \cite{dmt}} \\
\multicolumn{6}{l}{$^{b)}$ Calculated with $N=40$,
$R^2=1.65$, $\xi_1=$}\\
\multicolumn{6}{l}{$\;\;$
$12.3$ cm$^{-1}$, $\xi_2=21.3$ cm$^{-1}$,
$\kappa_4=- 14.8$ cm$^{-1}$}\\
\multicolumn{6}{l}{$\;\;$
and $\zeta = -0.38$ cm$^{-1}$} \\
\multicolumn{6}{l}{$^{c)}$ From \cite{vibron2}} \\
\end{tabular}
\end{table}

\end{document}